\documentclass[psfig,times]{aa}  
\input{psfig.sty}
\input{times.sty}
\begin{document}
\def\Mr{${\rm M}_{\rm R}$ }
\def\mr{${\rm m}_{\rm R}$ }
\def\re{${\rm r}_{\rm e}$ }
\thesaurus{ (03.13.2;     
	     11.01.2;     
	     11.02.1;     
	     11.06.2;     
	     11.09.2;     
	     11.16.1)}	  
\title{High$-$resolution imaging of Einstein Slew Survey BL Lacertae 
objects
\thanks{Based on observations made with the Nordic Optical
Telescope, operated on the island of La Palma, jointly by Denmark, Finland,
Iceland, Norway, and Sweden, in the Spanish Observatorio del Roque
de los Muchachos of the Institute de Astrofisica de Canarias.}
}
\author{J. Heidt\inst{1}, K. Nilsson\inst{2}, A. Sillanp\"a\"a\inst{2}, 
L.O. Takalo\inst{2}, and T. Pursimo\inst{2} }
\institute{Landessternwarte Heidelberg, K\"onigstuhl,
69117 Heidelberg, Germany
\and 
Tuorla Observatory, FIN$-$21500 Piikki\"o, Finland}
\offprints{\ \protect\\
 J.~Heidt,~E$-$mail:~jheidt@lsw.uni$-$heidelberg.de}
\date{Received $<$ date $>$; accepted $<$ date $>$ }
\maketitle
\markboth{J. Heidt et al.: High$-$resolution imaging of Slew BL Lac objects}{}
\begin{abstract}

High$-$resolution images of 7 newly identified BL Lac objects 
(among them one BL Lac candidate) at z $\leq$ 0.2 
from the Einstein Slew Survey are presented for the first time. 
In all cases we were able to 
resolve the host galaxy. Our 2$-$dimensional analysis of the host galaxies
shows that all these BL Lac objects are embedded in elliptical galaxies
with an average \Mr = $-$23.1 and \re = 10 kpc. One BL Lac might have
both a bulge and an underlying disk. These results are similar
to those obtained for the hosts of other BL Lac objects.

We searched in our BL Lac objects for host galaxies, whose surface 
brightness distribution does not follow a pure de Vaucouleurs law and 
determined the statistical significance with numerical simulations. 
In two BL Lac objects (1ES 1255+244 and 1ES 1959+650) significant 
deviations were found.

The environments of the BL Lac objects are highly interesting. In at
least two (perhaps three) cases we found  evidence for interaction.
All BL Lac objects (except one) have at least 2 companions,
some of which are bright, within a projected distance of 60 kpc from
the BL Lac. In two cases we found 5 companions within 50 kpc. This
implies that gravitational interaction is potentially important  to the
BL Lac phenomenon at least in these sources.


\end{abstract}

\section{Introduction}

BL Lac objects, characterized by strong variability from the radio 
up to the gamma regime, variable polarization in the radio and optical 
domain as well as weak or absent emission lines in their spectra, are 
the most enigmatic active galactic nuclei (AGN). Due to their extreme 
properties it is nowadays believed that their energy output is dominated
by Doppler$-$boosted synchrotron radiation arising from a relativistic
jet, which is seen in a  very small inclination angle (Urry \& Padovani 1995).
As a consequence observing their host galaxies and close environment
is a challenge for observers, since the light of the relativistic
jet in many cases overwhelms the light from the host galaxy.

However, such studies are very important.
By comparing isotropic properties of the individual subclasses of radio$-$loud 
AGN, the ``Unified Scheme'' can be tested.
For example, one would expect to find similar host galaxies for
BL Lac objects and their putative parents,
the low$-$luminosity Fanaroff$-$Riley I radio galaxies. 
Comparing the properties of
BL Lac host galaxies with ``normal'' (inactive) galaxies may give clues
as to why some galaxies harbor an AGN and (perhaps) others not.
Finally, they are one key to search for 
indications that the BL Lac phenomenon is related to merging/interaction
processes of galaxies (feeding the monster) as seems to be the case for 
quasars (e.g. Hutchings \& Neff 1992, Bahcall et al. 1995). 

In recent years much effort has gone into studies of the host galaxies 
of BL Lac objects using ground$-$based telescopes, leading to
a rapidly increasing number of resolved host galaxies up to redshifts
of $\sim 0.7$. 
Most of them (perhaps all) are hosted by luminous elliptical 
galaxies (${\rm M}_{\rm R} \sim -23.5$; e.g. Ulrich 1989, 
Abraham et al. 1991; Stickel et al. 1993; 
Falomo 1996; Wurtz et al. 1996). Observations with the HST, providing 
superior resolution, gave essentially the same results (Falomo et al. 1997,
Jannuzi et al. 1997).

The immediate environment has so far not been studied in detail. Falomo et al. 
(1990) and Falomo (1996) noted the high frequency of 
close ($<$ 40 kpc), mostly faint (${\rm M}_{\rm R} < -20$ if at the same 
redshift) companions among $\sim$ 20 observed BL Lac objects. 
For some BL Lac objects, it could be shown that
their (bright) companions are either at a similar redshift or that they are
physically associated (e.g. Stickel et al. 1993; Pesce et al. 
1994, 1995). Finally there are a few cases, where signs 
of interaction and physical association have been observed through 
imaging and spectroscopy  (e.g. Falomo et al. 1995).

In this paper we report the first high$-$resolution images of 7 newly
identified BL Lac objects (among them one BL Lac candidate) taken from the 
Einstein Slew Survey sample (Perlman et al. 1996; hereafter P96). 
Their X$-$ray to radio flux ratios log ${\rm f}_{\rm x}/{\rm f}_{\rm r}$ 
are $> -11$, which defines them as high$-$energy cutoff BL Lac objects
(HBL, Giommi \& Padovani 1994, Padovani \& Giommi 1995).
Due to their low redshifts (z $\leq 0.2$) we were able to carry out a
fully 2$-$dimensional analysis of their host galaxies. This allowed us not only
to study the morphology of the host galaxies in detail, but also to
investigate their close ($<$ 50 kpc) environment.

This paper is organized as follows. In Sect. 2 the observations and 
data reduction are summarized followed by a description of the data analysis
in Sect. 3. In Sect. 4  we give a short overview of the results and describe
than each object individually. The properties of the hosts as well
as the environment are discussed in Sect. 5. 
Finally, we summarize in Sect. 6.
Throughout the paper ${\rm H}_{\rm 0} =$ 50\ km\ 
${\rm s}^{\rm -1}\ {\rm Mpc}^{\rm -1}$ 
and ${\rm q}_{\rm 0} = 0$ is assumed.

\section{Observations and data reduction}

The observations were carried out with the Nordic Optical Telescope
on the nights July 8/9, 10/11 and 12/13 1996. A 1k CCD (scale 0.176\arcsec/pixel)
and a R filter was used throughout. For most of the time we enjoyed 
excellent seeing conditions ranging from 0.65\arcsec\ up to 1.1\arcsec\ FWHM. 
The exposure times
varied between 120s and 1800s depending on the brightness of the 
objects and the requirement not to saturate the BL Lac. The nights were 
photometric, standard stars from Landolt (1983) were frequently observed 
during each night to set the zero point.

The data were reduced (debiased, flatfielded using twilight flatfields),
cleaned of cosmics, aligned and coadded. Dark current was proven to be 
negligible. A journal of the observations along with the targets, their 
redshift, the total integration time, the measured FWHM and the 
conversion from angular to linear scale is given in Table 1.
The redshifts are from P96. For 1ES 1745+504 and 1ES
2037+521 (the former being a BL Lac candidate) no redshift is known yet.
Based on the results of our analysis we assumed z = 0.45 and 0.05,
respectively (see Sect. 4.1). The resulting redshift$-$dependent parameters
are given in parentheses.

\begin{table}
\caption[]{Journal of the observations.}
\begin{tabular}{ccrcc}
\hline
 & & & & \\
Object & z & Exp. [s] & FWHM [\arcsec] & kpc/1\arcsec \\
 & & & & \\
\hline
 & & & & \\
1ES 1255+244 & 0.141  &  900 & 0.67 & 3.41 \\
1ES 1440+122 & 0.162  & 1200 & 0.65 & 3.81 \\
1ES 1745+504 & (0.45) & 1800 & 1.11 & (7.71) \\
1ES 1853+671 & 0.212  &  720 & 0.88 & 4.69 \\
1ES 1959+650 & 0.048  &  990 & 0.67 & 1.32 \\
1ES 2037+521 & (0.05) & 1500 & 0.97 & (1.37)\\
1ES 2326+174 & 0.213  &  960 & 0.79 & 4.71 \\
 & & & &\\
\hline
\end{tabular}
\end{table}

\section{Data analysis}

In order to model the BL Lac objects and their hosts we applied a fully
2$-$dimensional fitting procedure developed at Tuorla Observatory to the
images. After masking close companions, faint features and projected
stars we fitted 6 different models (three galaxy models with and without
a nuclear point source) to the observed light distribution. The surface 
brightness I(r) of the galaxy at a distance r from its center was 
obtained from 

\begin{equation}
{\rm I}({\rm r}) = {\rm I}_{\rm e} {\rm dex} \left\{ -{\rm\bf b}_{\bf \beta} 
\left[ \left( \frac{{\rm r}}{{\rm r}_{\rm e}} \right) ^{\beta}
-1 \right]
\right\}
\end{equation}

where \re is the effective (half$-$light) radius of the galaxy, 
$\beta$ is a shape parameter and ${\rm b}_{\beta}$ is a $\beta-$dependent 
constant (Caon et al. 1993). For our galaxy models we chose $\beta$ = 
0.25 (de Vaucouleurs profile), $\beta$ = 1.0 (exponential disk) or
left $\beta$ as a free parameter. All galaxy models were convolved with 
the observed PSF.
The nuclear point source was modeled 
with a scaled PSF, obtained by averaging several well exposed stars on
the same frame as the BL Lac. We carefully checked for any PSF variations 
across the frames, but did not find any, except in the central
regions ($\leq$ 2 pixel).

Our models had 6 to 10 free parameters depending on the model: the (x,y)
center and magnitude of the core and the (x,y) center, magnitude,
effective radius, ellipticity, position angle and optionally the
$\beta-$parameter of the galaxy. The parameters were adjusted using
an iterative Levenberg$-$Marquardt loop to find the minimum $\chi^2$
between the model and the observed image. 

The analysis of 1ES 1440+122 was very difficult due to the presence of
a bright, close (projected distance $\sim$ 2.5\arcsec) companion.
To overcome this, we used an iterative fitting procedure
by fitting one of the two components first, subtracting it from the image, 
fitting the second, subtracting it from the original image, fitting 
the first etc. The iterations were stopped and final parameters adopted
when the fitted models ceased to improve significantly.

Finally, the models for the BL Lac core and their hosts as well
as for the bright companion of 1ES 1440+122
were subtracted from the images and aperture photometry performed 
of non$-$stellar objects (presumably galaxies) within a radius of 
100 kpc from the BL Lac.
Absolute magnitudes of the hosts were calculated using K$-$corrections
adopted from Bruzual (1983), the galactic extinction was estimated from
Burstein \& Heiles (1982). For the absolute photometry of 
the companions we used the (estimated) redshift of the BL Lac.

\begin{table*}[t]
\caption[]{Properties of the host galaxies of our BL Lac objects.}
\begin{tabular}{ccccrrcrrcrr}
\hline
 & & & & & & & & & & & \\
(1) & (2) & (3) & (4) & (5) & (6) & (7) & (8) & (9) & (10) & (11) & (12)\\ 
Object & 
${\rm m}_{\rm core}$  & ${\rm m}_{\rm host}$  & F & ${\rm r}_{\rm e}$ [``] & 
${\rm r}_{\rm e}$ [kpc] & $\epsilon$ & PA & ${\rm M}_{\rm host}$ & $\beta$ & 
$\Delta {\rm m}_{\rm core}$ & $\Delta {\rm m}_{\rm host}$\\
 & & & & & & & & & & & \\
\hline
 & & & & & & & & & & & \\
1ES 1255+244 & 17.9 & 16.7 & 0.75 &  2.1 &  7.2   & 0.10 &   5 & $-$23.2  
& 0.18 & 0.15 & $-$0.16 \\
1ES 1440+122 & 17.6 & 17.2 & 0.59 &  2.3 &  8.8   & 0.18 & 170 & $-$23.0  
& 0.21 & 0.03 & $-$0.09 \\
1ES 1745+504 & 20.5 & 19.8 & 0.66 &  1.1 & (8.5)  & 0.17 & 165 & ($-$23.2)
& 0.28 &  $-$0.04 & 0.06 \\
1ES 1853+671 & 19.2 & 18.1 & 0.73 &  2.0 &  9.4   & 0.12 & 155 & $-$22.9  
& 0.32 &  $-$0.08 & 0.10 \\
1ES 1959+650 & 15.2 & 14.8 & 0.59 &  9.5 & 12.5   & 0.21 &  95 & $-$23.0  
& 0.41 &  $-$0.05 & 0.28 \\
1ES 2037+521 & 19.3 & 15.9 & 0.96 &  9.0 & (12.3) & 0.19 & 120 & ($-$23.2)
& 0.20 & 0.25 & $-$0.20 \\
1ES 2326+174 & 18.3 & 17.5 & 0.68 &  1.8 &  8.5   & 0.17 &  95 & $-$23.4  
& 0.17 & 0.17 & $-$0.18 \\
 & & & & & & & & & & & \\
\hline
\end{tabular}
\end{table*}

\section{Results}

Although the presence of close companions (some of which are bright),
stars or faint associated features  made the analysis difficult for most
of the sources,
in every case a de Vaucouleurs model for the host galaxy gave a much better 
fit to the data than an exponential disk model. Leaving $\beta$ as free 
parameter improved our fits as compared to the fits with $\beta$ = 0.25
only marginally, with the exception of 1ES 1959+650 and 1ES
1255+244 and perhaps 1ES 2037+521. For 1ES 1959+650 we carried out further
modeling, which will be described in detail in Sect. 4.1.
We also estimated the significance of a deviation of $\beta$ from 0.25
by numerical simulations, these are described in Sect. 5.1.
In what follows, we will refer to the results of our fits with $\beta$
= 0.25.

The results of our modeling are summarized in Table 2. 
Column 1 gives the object name, 
in columns 2 and 3 the magnitudes for the core and the host as derived
from the model and in column 4 the host to total flux ratios F are presented.
In columns 5 and 6 the half$-$light radii along the major axes 
in arcsec and kpc are shown,
followed by the ellipticity in column 7 and the position angle PA
(counted counterclockwise from north) in column 8. In column 9
the calculated absolute magnitudes are given. Finally, we display in
columns 10$-$12 the results with $\beta$ as free parameter and the
resulting changes of the core and host magnitudes as compared to
the fits with $\beta$ = 0.25. As in Table 1, we assumed z = 0.45
for 1ES 1745+504 and z = 0.05 for 1ES 2037+521. Their redshift$-$dependent
parameters are given in parenthesis.

In Figs. 1a$-$f and 2b$-$d we show the central region around 
each BL Lac before and after the subtraction of our model (host with
$\beta$ = 0.25+core except 1ES 1959+650, where the core and 
both the bulge and disk was 
subtracted). Although the residuals close to the center of the fitted objects 
sometimes look strong, they are in all cases fairly small compared to the
total signal ($<$ 5\%). 

\subsection{Results for individual objects}

{\bf 1ES 1255+244}: The redshift of this source is z = 0.141 (P96). 
Our fitting procedure gives \Mr = $-$ 23.2 
with \re = 7.2 kpc and small ellipticity ($\epsilon$ = 0.1). 
The model fit is excellent as can be seen in Fig. 1a. 

There are 2 brighter
galaxies within 15\arcsec\  from the BL Lac (A and B in Fig. 1a),
with \mr = 20.4 and 20.6, respectively. Thus they would be fainter than
\Mr = $-$20, if at z = 0.141. The closest companion (A) is at a projected 
distance of 26 kpc. 
A relatively bright (\mr = 17.9) disk$-$type edge$-$on galaxy can be seen
23\arcsec\ (80 kpc) to the north (C), which would have \Mr = $-$22 
at z = 0.141. Superimposed are a number of faint galaxies 
(e.g. to the west of the BL Lac).

{\bf 1ES 1440+122}: This is the most intriguing BL Lac object of
our small sample. Its redshift is z = 0.162 (Schachter et al. 1993).
The most remarkable feature is a very close, luminous companion 2.5\arcsec\
(projected distance 9.5 kpc) to the west (Fig. 2b). As described in section 
3.2 we used an iterative fitting procedure to derive the parameters for 
the host galaxy of the BL Lac and for the close companion. The host galaxy 
is of typical brightness (\Mr = $-$23.0), size (\re = 8.8 kpc) and  
moderate ellipticity ($\epsilon$ = 0.18). The decentering
(core vs. host centroids for the BL Lac) is 
negligible ($\sim 0.02$\arcsec). The companion is best fit by 
an early$-$type galaxy of similar brightness (\Mr = $-$23.2) and 
size (\re = 8.4 kpc), but of higher ellipticity ($\epsilon = 0.3$). 

The residuals after the subtraction of our model for the BL Lac and companion 
are complex (Figs. 2c, d). 
Very close to the center of both fitted objects there are doughnut$-$like 
residuals, which are larger for the companion than for the BL Lac.
Roughly along the major axis of the companion (north$-$south)
there is additional diffuse emission left on both sides in the
outer part of the galaxy.
On top of the diffuse emission in the north is a low
surface brightness feature $\sim$ 5\arcsec\ from the center of the companion
(already visible on the original frame). 
This feature could be a projected faint galaxy
(B in Fig. 2d), although we can not rule out
that it is part of the BL Lac or the bright companion. 
Whereas the residuals close to the center of both objects
are artifacts due to the fitting procedure, the outer residuals
for the companion $-$  in spite of feature B $-$ may be intrinsic to the source.

In order to test this, we carried out an isophotal analysis of 
the companion (after subtraction of our model for the BL Lac core + host) by
fitting ellipses to the images according to the method outlined
by Bender \& M\"ollenhoff (1988). This method provides the azimuthally 
averaged surface brightness,
ellipticity, position angle, the centers of the ellipses as well as the 
Fourier coefficients as a function of the semi$-$major radius {\it a} and 
semi$-$minor radius {\it b}.
Our isophotal analysis of the companion shows clear indications 
of an isophote twist. Whereas $\epsilon$  grows from 0.2 at r = 2\arcsec\ to
0.45 at 4.5\arcsec\ and beyond, PA decreases from $\sim 30^{\rm o}$ at 
2\arcsec\ to $\sim 10^{\rm o}$ at 4.5\arcsec. 

\begin{figure*}
\centerline{\hbox{
\hspace*{.5cm}
}}
\vspace*{.5cm}
\centerline{\hbox{
\hspace*{.5cm}
}}
\caption [] {Images of the BL Lac objects observed. The left panel
shows the central part of the original frames, the right panel the 
central part of the frames after
subtraction of the BL Lac (convolved de Vaucouleurs model galaxy and scaled PSF,
except 1ES 1959+650, where the model galaxy consists of a bulge and a disk). 
Galaxies discussed in the text are labeled, stars are labeled with ``S''.
North is up, east to the left. 
The grey scale is logarithmic in order to enhance low surface brightness 
features. a) 1ES 1255+244, 60\arcsec$\times$60\arcsec\ (200$\times$200 kpc), 
b) 1ES 1745+504, 34\arcsec$\times$34\arcsec\ (150$\times$150 kpc),
}
\end{figure*}

\setcounter{figure}{0}

\begin{figure*}
\centerline{\hbox{
\hspace*{.5cm}
}}
\vspace*{.5cm}
\centerline{\hbox{
\hspace*{.5cm}
}}
\caption [] {$-$ continued. 
c) 1ES 1853+671, 18\arcsec$\times$18\arcsec\ (85$\times$85 kpc),
d) 1ES 1959+650, 42\arcsec$\times$42\arcsec\ (55$\times$55 kpc),
}
\end{figure*}

\setcounter{figure}{0}

\begin{figure*}
\vspace*{.5cm}
\centerline{\hbox{
\hspace*{.5cm}
}}
\vspace*{.5cm}
\centerline{\hbox{
\hspace*{.5cm}
}}
\caption [] {$-$ continued. 
e) 1ES 2037+521, 24\arcsec$\times$24\arcsec\ (33$\times$33 kpc),
f) 1ES 2326+174, 44\arcsec$\times$44\arcsec\ (200$\times$200 kpc).
}
\end{figure*}

\begin{figure*}
\centerline{\hbox{
\hspace*{.5cm}
}}
\vspace*{.5cm}
\centerline{\hbox{
\hspace*{.5cm}
}}
\caption [] {a) Large scale environment of 1ES 1440+122 in order to show the 
galaxy enhancement close to the BL Lac. Field is 126\arcsec$\times$126\arcsec\ 
(430$\times$430 kpc). 
b) Central 18\arcsec$\times$18\arcsec\ (69$\times$69 kpc). The BL Lac
is the left object of the pair of galaxies. c) Same as b) with the
model for the BL Lac (core + host) subtracted to show the close companion.
d) Same as b) with the model for the BL Lac and the close companion subtracted.
North is up, east to the left. The grey scale is logarithmic except a), where
a linear scale was used.}
\end{figure*}

If we assume that the BL Lac and the companion are at the same redshift,
our observations and the isophotal analysis suggests that we observe
an interacting pair of very bright elliptical galaxies, one of which hosts an 
active nucleus. We note that the redshift of 1ES 1440+122
could be contaminated by its close companion. Therefore to verify
our scenario redshifts of both galaxies
are required.

The environment of 1ES 1440+122 is very interesting. 
The BL Lac is surrounded by $\sim$ 20 galaxies (\mr $\sim$
20$-$22), suggesting that 1ES 1440+122 is located in a group or small cluster
of galaxies. Most of the galaxies are within 200 kpc of the BL Lac
(Fig. 2a). We note that 5 galaxies (including the bright companion) 
are found within 8\arcsec\ (30 kpc, A$-$E in Figs. 2c, d).

{\bf 1ES 1745+504}: This source is the BL Lac candidate from our
Slew Survey subset. Its classification is still uncertain 
and no redshift is available (Perlman, priv. communication).
Although our image of this source was taken under the worst seeing
conditions (FWHM $\sim$ 1.1\arcsec), the host galaxy is
clearly resolved. Moreover, there is a faint feature towards the 
north (A in Fig. 1b). After proper masking this feature, 
the host can be fitted very well by an early$-$type galaxy with 
\mr = 19.8 and \re = 1.1\arcsec\ (Fig. 1b). 

Under the assumption that the host of 1ES 1745+504 has similar properties
as the hosts of other BL Lac objects, we can try to estimate a redshift for 
1ES 1745+504. The average \Mr of the hosts of our BL Lac objects with
firm redshift is $-$23.1 and the average \re is 9.3 kpc (see table 2).
If we adjust the redshift of 1ES 1745+504 such that we derive similar values,
a good guess would be z = 0.45. In that case we would derive
\Mr = $-$ 23.2 and \re = 8.5 kpc (including K$-$correction and galactic 
extinction). For z = 0.4 \Mr would be $-$22.9, \re = 7.9 kpc and for
z = 0.5 \Mr would be $-$23.6 and \re = 9 kpc.
We will use z = 0.45 hereafter.

After subtraction of our model from the image, the feature to the north 
appears to be resolved, round and well separated from the BL Lac.
Therefore, we consider it rather to be a galaxy  
than a tidal tail or a merger remnant possibly associated with the BL Lac. 
The integrated magnitude of this system is \mr $\sim$ 23 at a projected 
distance of 3.5\arcsec.

Within 15\arcsec (110 kpc at z = 0.45) of the BL Lac 5 more companion galaxies
are present (B$-$F in Fig. 1b), ranging in brightness from \mr = 
17.9 up to 21.1. This suggests that they may form together with the
BL Lac a small group. 

{\bf 1ES 1853+673}: In view of the short integration time as opposed
to its redshift (z = 0.212, P96) we secured a good image of this source,
which shows interesting features. The host galaxy is clearly resolved,
and again, a close, resolved companion 2\arcsec\ to the northwest 
(9.4 kpc at z = 0.212)
can be seen. Moreover, there is a low surface$-$brightness feature
associated with the companion which extends $\sim$ 7\arcsec\
to the northwest (the companion is labeled A and the feature indicated
with an arrow in Fig. 1c). The optical appearance is striking and 
suggestive of a tidal tail resulting either from an interaction between both 
galaxies or a merging process, where material was stripped off  
the companion.

After we had properly masked the companion and the ``tidal tail''
we were able to make a very good fit to the system. The host galaxy is
very round ($\epsilon = 0.12$) with \Mr = $-$22.9 and \re = 9.4 kpc. 
Photometry of the companion gives \mr $\sim$ 21.8,
which would imply \Mr = $-$19.2 at z = 0.212. Still the ``tidal tail'' is 
present, but due to its low surface brightness, we did not attempt to
estimate a magnitude of that feature.

There is only one more galaxy within 11\arcsec\ (50 kpc) of the BL Lac 
(B, \mr = 23.1). 

{\bf 1ES 1959+650}: This is the closest BL Lac of our targets (z = 0.048, P96, 
Schachter et al. 1993). We secured a very good image under excellent seeing
conditions (0.67\arcsec\ FWHM). The modeling of the host was slightly 
difficult due to a presence of a bright star 10\arcsec\ to the northwest.

None of our 6 different models fitted to the image gave satisfactorily 
results. Alternatively, we fitted a 3 component 
model consisting of a core, a disk and a bulge (with $\epsilon$ free and
PA equal for disk and bulge). In all cases, we had residuals close
to the core as well as in the outer parts of the host galaxy left.

A careful inspection of our image of 1ES 1959+650 before and after
the subtraction of our models showed always a highly interesting feature
(a hint of this feature is already visible on 
the original frame). Approximately 1\arcsec\ to the north of the 
center of 1ES 1959+650 there
is an absorption feature roughly oriented along the major$-$axis
of the host galaxy in E$-$W direction, suggestive
of a dust lane (indicated in Fig. 1d with an arrow).
If this could be confirmed, we would have found the first
BL Lac hosted by a dust$-$lane elliptical galaxy with an underlying disk!

We masked the absorption feature and repeated the fitting procedure as
described above. The fit with a de Vaucouleurs model gave
\mr = 14.77 with \re = 11.3 kpc and small ellipticity ($\epsilon$ = 0.20).
However, best fits are obtained with either the model with core and 
$\beta$ = 0.41 or with core and bulge and disk. For the latter we obtained
\mr = 15.2 for the core, \mr = 15.0 and 
\re = 10.2\arcsec\ (13.5 kpc) for the bulge and \mr = 16.7
and \re = 4.8\arcsec\ (6.3 kpc) for the disk. The ellipticities are
0.01 for the bulge and 0.47 for the disk implying an inclination
of $\sim 58^{\circ}$. The disk$-$to$-$bulge ratio is $\sim$ 0.22 and 
\Mr of the whole galaxy (disk and bulge) = $-$23.0.

Although we masked the absorption feature carefully, the residuals close 
to the center and in the outer parts of the host galaxy are
still strong as compared to the other sources modeled.
For the former we checked the quality of our PSF by subtracting a scaled
version of it from different stars on different locations on the CCD frame 
and found always a very good match, except the inner 2 pixels. 
The residuals close to 
the center could still be a remnant of imperfect masking.

For the residuals in the outer part of the host galaxy, the
absorption feature can not be the reason alone.
In fact, our isophotal analysis (absorption feature masked) 
shows that in 1ES 1959+650 an isophote
twist is occurring. The PA changes from $\sim 95^{\circ}$ at r = 3\arcsec\
up to $140^{\circ}$ at r = 15\arcsec\ and $\epsilon$ changes from
0.2 to 0.

Basically, two scenarios could be responsible for our residuals and
the observed change of PA and $\epsilon$.
Either the disk and bulge have not only different ellipticities, but
also different PA, or the whole galaxy ($\beta$ = 0.41) or
at least the bulge (in the 3 component fit) is triaxial (Benacchio \& Galletta,
1980).

Isophote twists can also be produced by interaction between galaxies 
(Madejsky, 1990). There are only three brighter galaxies on the full frame
24\arcsec, 37\arcsec\ and 55\arcsec\ to the southeast, south and northeast,
respectively (31, 49 and 73 kpc at z = 0.048 with \mr = 20.0, 18.4 and 20.1,
the closest labeled A in Fig. 1d). Since none of them shows signs of 
interaction with 1ES 1959+650, we consider this scenario as unlikely.

{\bf 1ES 2037+521}: This source was listed as BL Lac candidate in P96.
Meanwhile it is confirmed as BL Lac spectroscopically, but still
no redshift is known (Perlman, priv. communication). 
Due to its low galactic latitude (b~$\approx 7^{\circ}$),
the image is crowded with stars. The host galaxy is well resolved,
but unfortunately a bright star 1.9\arcsec\ northeast of the center of 
1ES 2037+521 made the analysis difficult.

In spite of the difficulty with the bright, closeby star (which we subtracted
before fitting) our model fits 
very well the observed light distribution (Fig. 1e). 
The host has \mr = 15.9 and \re = 9.0\arcsec, 
the core is $\sim$ 3.5 mag fainter (\mr = 19.3). 
Thus the light of the host galaxy dominates almost entirely  the system 
($\sim$ 96\% from the total flux of 1ES 2037+521 came from the host 
during our observations).

In order to estimate the redshift of 1ES 2037+521, we adopt a similar
strategy as in the case of 1ES 1745+504. We adjust the redshift 
in such a way that
the resulting redshift dependent parameters for the host are similar to those 
for our BL Lac objects with firm redshift.
Here, however, is an additional difficulty due to
the low galactic latitude of the object, which made the assumption of
the galactic absorption very uncertain. Based on Burstein \& Heiles (1982)
we assume A(R) = 1.65 mag. The error may be as large as 0.3 mag.

Therefore a good guess for the redshift of 1ES 2037+521 would be
z = 0.05. We would derive \Mr = $-$23.2 and \re = 12.3 kpc 
(for z = 0.1 \Mr would be $-$24.8 and \re = 23 kpc, which would make
the host of 1ES 2037+521 among the brightest and largest known).

Similarly to BL Lac itself, which is also a low galactic latitude source,
there are only a few galaxies with \mr = 20$-$22 
on the full frame. No obvious companions within 30\arcsec\ are present.

{\bf 1ES 2326+174}: This is the object with the highest redshift in our sample
(z = 0.213, P96). Superimposed onto the outer parts of the host there are 
three faint galaxies (\mr $\approx$ 22.5$-$23.5)
3.4\arcsec, 3.1\arcsec\ and 6.6\arcsec\ to the west, south and east, respectively 
(16.0, 14.1 and 31.2 kpc if at z = 0.213, A$-$C in Fig. 1f).

The host can be well fitted  by a de Vaucouleurs law with
\Mr = $-$23.4 and \re = 8.5 kpc.
After subtracting the model for the BL Lac, the three close companions are
well visible, with no obvious sign of interaction.  
Within 20\arcsec\ ($\approx$ 100 kpc at z = 0.213) of the BL Lac there are 
three brighter galaxies (\mr = 19.5$-$20.5, D$-$F in Fig. 1f) 
which may together with the BL Lac form a small group.

\section{Discussion}

\subsection{Properties of the host galaxies}

Since we were able to carry out the observations of our sources under good,
partly excellent seeing conditions and since most (perhaps all) of our
sources are at redshifts $\leq$ 0.2, we were able resolve the host galaxies
in all cases. This allowed us to apply a 2$-$dimensional fitting 
procedure to the images. In all cases, except one (1ES 1959+650), the
host galaxy can best be 
modeled by a shape parameter $\beta$ close to the canonical de Vaucouleurs 
value. In none of the cases did a disk$-$type host galaxy give an 
acceptable fit.

The results are well in the range found for other BL Lac objects
by various groups. 
Falomo (1996) found for his BL Lac objects with z $<$ 0.2 a
$\langle$ \Mr $\rangle = - 23.4\pm 0.7$, while Wurtz et al. (1996) found
$\langle$ \Mr $\rangle = - 23.7\pm 0.6 $ for z $<$ 0.2 BL Lac objects.
 We determined $\langle$ \Mr $\rangle = - 23.1\pm 0.2$ 
for the 5 BL Lac objects with firm redshift which is in excellent agreement.
For the half$-$light radii we derived 
$\langle$ \re $\rangle = 9.3\pm2$ kpc, which is similar to most z $<$ 0.2
BL Lac objects observed by Wurtz et al. (1996), Falomo (1996) and 
Abraham et al. (1991) but not as large as 51 kpc determined for
PKS 0548$-$322 (Falomo et al. 1995). 

We found ellipticities ($\langle \epsilon \rangle = 
0.16\pm0.05$) which are similar to those obtained from ground
(Falomo 1996; Abraham et al. 1991) and with HST (Falomo et al. 1997;
Jannuzi et al. 1997).
The offsets between the core and host centroids are small ($<$0.1\arcsec).
In all cases the light from the host galaxy dominates the system 
from 59\% in 1ES 1440+122 and 1ES 1959+650 up to 96\% in 1ES 2037+521.
This is typical for HBL (see e.g. Wurtz et al. 1996).

The residuals after subtraction of our model (core + convolved host)
are in all cases fairly small ($<$ 5\% of the total light), 
except for 1ES 1959+650. These 6 sources have $\beta$ close to 0.25, 
which shows that in these cases a de Vaucouleurs law is a good
representation of the light distribution from the host galaxy. 

One might ask, however, if the deviations from a pure de Vaucouleurs law 
observed here are significant or simply due to noise or systematic
effects. An apparent deviation could be caused by at least 1) systematic 
errors in the fitting program, 2) photon and readout noise in the
images, 3) incorrect background subtraction or 4) incorrect
PSF. We have performed simulations to study the effect of 1) - 3)
to the fitted $\beta$ values. 
The effects of PSF errors are not accounted for in our simulations, but
we expect these effects to be small given the relatively small 
residuals in the model subtracted images. A simulation including
also the PSF effects would be very useful provided that they could be 
included in a realistic way. 

Thus we created $\sim$ 100 simulated images 
for each object taking the parameters for the core and host galaxy
from our de Vaucouleurs fits. The host galaxy obeyed strictly the de 
Vaucouleurs profile in each simulated image. The images were 
convolved with the PSF, and photon noise,
readout noise and a constant drawn from a Gaussian distribution with
zero mean were added. The constant represents the error in the sky 
subtraction and its standard deviation was determined for each
object from the sky measurement. Each simulated image was then fit
with 10 free parameters, i.e. by letting $\beta$ to be a free
parameter. Masking of images was identical to the actual fits. 

The first remark from the simulations is that in all 7 objects most of
the uncertainty
in $\beta$ (and m$_{\rm core}$, m$_{\rm host}$ and \re as well) comes from
the uncertainty in the sky level; photon and readout noise have less
effect on $\beta$ in the S/N conditions present in our images. 
Also, the fitted values were scattered symmetrically around their
true values showing that no bias is introduced by the fitting program.
The sensitivity to assumed sky level emphasizes the importance of
accurate sky subtraction if one wants to obtain correct host galaxy
parameters. 

Secondly, the final
$\beta$ values in the simulated images cluster around $\beta$ = 0.25
with a standard deviation that varies from object to object. The
distributions of $\beta$ are nearly Gaussian, so we calculated the 
standard deviation $\sigma_{\beta}$ and 99\% (3 $\sigma_{\beta}$) 
confidence intervals for each object. In two objects, 1ES 1255+244 
($\beta$ = 0.18, $\sigma_{\beta}$ = 0.02) and 1ES 1959+650 
($\beta$ = 0.41, $\sigma_{\beta}$ = 0.01) the fitted $\beta$ lies
outside the 99\% confidence interval. In 1ES 2037+521 ($\beta$ = 0.20,
$\sigma_{\beta}$ = 0.02) the fitted $\beta$ lies inside the 99\% but 
outside the 95\% interval. 
Thus we have evidence of $\beta$ significantly different from 0.25 in 
two objects and marginal evidence in one object. Because we have not 
included the PSF errors in our simulations, the confidence intervals
we calculate have to be regarded as lower limits. Thus 1ES 2037+521
might move inside the 95\% limits after including the PSF errors,
whereas the results for 1ES1255+244 and especially 1ES 1959+650 are 
more secure in this respect.

The results of our simulations may shed new light on the morphology 
of BL Lac host galaxies, although the amount of data is small. 
Most previous studies used either pure de Vaucouleurs or disk-type
fits to determine their nature. 
In their 2$-$dimensional analysis of BL Lac host galaxies from 
the 1 Jy sample Stickel et al. (1993) left $\beta$ as a free
parameter, but did not convolve the galaxy models. 
It is nevertheless remarkable that according to their result the host
galaxy of BL Lac itself can best be fitted by $\beta$ = 0.66 in the
R-band.
With the current availability of large telescopes offering excellent 
seeing conditions and sufficient resolution in combination with a dedicated
2$-$dimensional analysis, we can now determine the fraction of BL Lac objects,
whose host galaxies do not follow a pure de Vaucouleurs law. The data
obtained during the HST Snap survey on BL Lac objects are perfectly suited for
this kind of work.

Given the large residuals after subtraction of our model (convolved de
Vaucouleurs + core) and the results from our numerical simulations, 
1ES 1959+650 seems to be hosted by the most peculiar galaxy among the
BL Lac objects we observed.
Both a fit with $\beta$ = 0.41 and a fit with a bulge and a disk for
the host galaxy gave best results. At present we can not distinguish
which of our models represents the true nature of the morphology of the
host galaxy of 1ES 1959+650. Caon et al. (1993) found in their study of
a sample of elliptical and S0 galaxies a wide range of $\beta$, with
basically no difference between both galaxy types, whereas Capaccioli et al.
(1992) noted that most early$-$type galaxies have disks. 

So far, only one BL Lac object has been found to be hosted by an elliptical
galaxy with a disk (PKS 0548$-$322; Falomo et al. 1995), but in this case the
disk was very small. 

There are also claims that a few BL Lac objects might be hosted by
a disk$-$type galaxy (Abraham et al. 1992, Wurtz et al. 1997), some of them
are discussed controversary (e.g. PKS 1413+135, Wurtz et al. 1997 and 
references therein).

In 1ES 1959+650 we found also an
indication for a dust$-$lane, which would be the first ever seen in any
BL Lac host. This is not unusual for elliptical or
disk$-$type galaxies. 
M\"ollenhoff et al. (1992) found 21 out of 26 dust$-$lane elliptical galaxies
to be radio emitters, 6 out of their 7 objects with unresolved radio 
structure are major$-$axis dust$-$lane elliptical galaxies. 
Since  1ES 1959+650 appeared unresolved at 6cm (P96) our findings would
be consistent with their results. Unfortunately, this indication of a dust 
lane is based on measurements in one passband only. Multicolour images
(e.g. in B band, where the dust lane should show up clearly and in 
the NIR, where the dust lane should not show up at all)
are needed to verify the presence of a dust lane.

\subsection{Environment}

\subsubsection{Evidence for interaction}

So far only a couple of BL Lac objects have been reported to show clear 
evidence of interaction with other galaxies. 
In both, Ap Lib and 3C 371, a companion galaxy at projected distances
of 110 and 83 kpc, respectively, have been found. Whereas the former
was characterized by asymmetric isophotes elongated towards the 
companion, the latter shows a tidal tail connecting both galaxies
(Arp 1970; Stickel et al. 1993; Nilsson et al. 1997). 
Spectroscopy by Stickel et al. (1993) and Pesce et al. (1994)
found the companion galaxies at the same redshift as the BL Lac thus 
confirming their physical association.

Striking evidence for interaction has been found in
PKS 0548$-$322 (Falomo et al. 1995). A relatively bright
companion galaxy separated by 25 kpc from the BL Lac was detected, with 
signs of interaction present via extended low surface brightness 
emission. Unfortunately, no redshift for this galaxy is known yet.

A similar object to PKS 0548$-$322 is 1E 1415.6+2557. Halpern et al. (1986)
noted a close, bright companion separated by 5\arcsec\ (26 kpc in our cosmology)  
from the BL Lac. No redshift of the companion is known. Even more close,
a jet-like feature 3\arcsec\ (15 kpc) from the BL Lac was detected by Romanishin
(1992), who interpreted this feature as an optical jet. This is still 
under debate, e.g. Gladders et al. (1997) interpreted this feature rather
as an inclined spiral in projection against the BL Lac or a nearby companion
galaxy.

In two (perhaps three) of our BL Lac objects we found  evidence for
interaction. 1ES 1440+122 has a close early$-$type companion 
(projected distance $\sim$ 2.5\arcsec) with a brightness differing
only by 0.2mag from that of the BL Lac 
host. If both objects would be at the same redshift, their cores
would be separated by $\sim$ 9.5 kpc only! We found strong support for
the interactions hypothesis by our 
isophotal analysis, which shows a clear isophote twist in the companion
along with strongly varying ellipticity. The relatively undisturbed appearance 
of both galaxies suggests that they are currently approaching rather than 
having already undergone a close encounter. 

The other source showing evidence of interaction was 
1ES 1853+671. Similarly to 1ES 1440+122 we found a close companion
(projected distance $\sim$ 2\arcsec\ = 9.4 kpc at z = 0.212), 
but $\sim$ 3.7 mag fainter than the host of 1ES 1853+671.
In this case, however, there is a faint extension from the companion 
extending $\sim$ 7\arcsec\ to the opposite side of the BL Lac. 
This feature is very similar to tidal tails
in interacting galaxies and a similar scenario is suggested here.
Contrary to 1ES 1440+122, the optical appearance suggests, that
the objects already had one close encounter, during which material was 
stripped off the companion. 
This system is similar to PKS 0548$-$322 and its close companion, 
but in the latter case the projected distance is 25 kpc.

Another source, which might be influenced by interaction is 1ES 1745+504,
although the evidence is weak. This source has a faint feature to
the north. After the subtraction of the model for the BL Lac, this
feature seems to be quite round and (perhaps) separated. Thus we
believe that this is rather a galaxy well separated from the BL Lac 
than an interacting system.

In summary, although this is just now tentative, our results imply that 
tidal interactions are potentially important to the BL Lac phenomenon at
least in some sources. 
1ES 1440+122 and 1ES 1853+671 may represent the extremes of
this phenomenon (approaching, but already very close galaxies for the former
and undergone interaction for the latter). Clearly, this issue
has to be tested by very deep high$-$resolution imaging using HST and
by spectroscopy from ground using large telescopes.

\subsubsection{Close companions}

Already Falomo et al. (1990) noted close ($<$5\arcsec), faint ($>21$mag)
emission features around a couple of BL Lac objects. Since then a few 
studies related to this subject have been carried through both, imaging
and spectroscopy (e.g. Stickel et al. 1993; Pesce et al. 
1994, 1995; Falomo et al. 1995; Falomo 1996; Jannuzi et al. 1997). 
Whereas the imaging data allowed one to identify close companions for 
almost all 
of the objects, their physical association could only be confirmed for some 
of them. This is mainly due to their faintness as compared to the BL Lac and 
their close separation, which makes it hard to measure their redshifts
even with large telescopes.

In all of our BL Lac objects $-$ except 1ES 2037+521 $-$ 
we found at least 2 companions within a projected distance of $\sim$ 60 kpc.
Most companions were found around 1ES 1440+122 and 1ES 1745+504
(5 within 30 kpc and 6 within 60 kpc, respectively). 1ES 2326+174 has
6 companions within 100 kpc, three of them within 30 kpc projected distance.
They span a wide range of brightnesses, between \mr = 17 and 23.5.
Whereas some of the companions are most likely projected along the line
of sight to the BL Lac (e.g. in 1ES 1745+504), and the faintest ones 
background galaxies, most of them could well
be at the same redshift as the BL Lac, or at least members of the 
putative cluster surrounding the BL Lac. 

These observations (and those reported in Sect. 5.2.1) 
point to the fact that interaction might be 
related to the BL Lac phenomenon as seems to be the case for QSOs 
(e.g. Hutchings \& 
Neff 1992; Bahcall et al. 1995; Miller 1998). That this is generally
the case, has not yet been convincingly demonstrated. 
One possible test would be a study of a well defined, low 
redshift sample of BL Lac objects, FR I radio galaxies (the assumed parent 
population of BL Lac objects) {\it and} a control sample of ``normal'' 
galaxies. The samples must well be matched in redshift and
luminosity space. Since both, imaging and spectroscopy is needed for such a
program (imaging for the number and luminosity distribution of the
companions, spectroscopy to check their physical association),
this would be a project requiring a lot of telescope time
on (very) large telescopes. And even when it could be shown that the samples
of active galaxies have e.g. on average more frequent and brighter
companions at the same redshift as the control sample of ``normal'' galaxies,
its not clear if every possible bias is taken into account. For example,
the proper selection of the control sample is very difficult, because it should
be chosen in such a way that it is not only well matched in redshift and 
luminosity distribution with the samples of active galaxies, but should
also have the same mix of large$-$scale surroundings (cluster environments).

\section{Summary}

We have presented the first images of 7 newly detected BL Lac objects 
(among them one BL Lac candidate) from the Einstein Slew Survey.
Since they are at low redshift and thanks to our excellent resolution,
we could resolve their hosts in all cases. This allowed us to analyze the
host galaxies, to study their properties by a 2$-$dimensional 
decomposition method and to investigate their environment.

The properties of the host galaxies are very similar to those 
derived for other BL Lac objects. They are luminous \Mr $\sim$ $-$23.1
and large \re $\sim$ 10 kpc elliptical galaxies. 
The host of 1ES 1959+650 is complex and may be either an elliptical galaxy with
$\beta$ = 0.41 or an elliptical ($\beta$ = 0.25) galaxy with an underlying 
disk. Based on the average properties of the BL Lac hosts with firm redshift,
we estimated a redshift of 0.45 for the BL Lac candidate 1ES 1745+504
and a redshift of 0.05 for 1ES 2037+521.

We searched for deviations of a pure de Vaucouleurs law ($\beta \neq$ 0.25)
in the host galaxies of our BL Lac objects and determined their 
statistical significance with numerical simulations. In two objects
(1ES 1255+244, 1ES 1959+650) we found significant deviations. This may shed 
new light on the properties of the host galaxies of BL Lac objects, but
due to the small number observations an analysis of a large sample
of BL Lac objects is required.

The environments of the BL Lac objects are highly interesting.
We found in at least two (perhaps three) cases evidence for interaction
(1ES 1440+122, 1ES 1853+671 and perhaps 1ES 1745+504). 
In all except one cases we found at least 2 companions within a 
projected distance of 60 kpc around the BL Lac. 
In two cases we found at least 5 within 50 kpc (1ES 1440+122, 1ES 1745+504). 
This may show that interaction is potentially important to  
BL Lac phenomenon at least in these sources. 

\acknowledgements{We thank the anonymous referee for his prompt
and constructive comments. Thanks to E. Perlman for an update of the 
current status 
of the Einstein Slew Survey sample of BL Lac objects.
We benefited from stimulating discussions on the morphology as well as the  
modeling of elliptical galaxies with J. Fried, C. M\"ollenhoff
and C. Scorza de Appl. This work was supported by the DFG
(Sonderforschungsbereich 328) and the Finnish Academy of Sciences.
TP acknowledges support from the Wihuri Foundation.}

\end{document}